\documentclass[%
  reprint,
 amsmath,amssymb,
 aps,
]{revtex4-2}

\usepackage{inputenc}
\usepackage{graphicx}
\usepackage{dcolumn}
\usepackage{bm}
\usepackage{xcolor}

\newcommand{\etal}{\textit{et al.\ }}

\begin{document}

\preprint{APS/123-QED}

\title{Quasiparticle self-consistent $GW$ band structures and phase transitions of LiAlO$_2$ in tetrahedrally and octahedrally coordinated structures}

\author{Phillip Popp}
\author{Walter R. L. Lambrecht}%
\email{walter.lambrecht@case.edu}
\affiliation{Department of Physics, Case Western Reserve University, 10900 Euclid Avenue, Cleveland, OH-44106-7079, USA
}%

\date{\today}

\begin{abstract}
A first-principles computational study is presented of various phases of LiAlO$_2$. The relative total energies and equations of state of the $\alpha$, $\beta$ and $\gamma$ phases are determined after structural relaxation of each phase. The $\beta$ and $\gamma$ tetrahedral phases are found to be very close in energy and lattice volume with the $\gamma$ phase having the lowest energy. The octahedral $\alpha$ phase is a high-pressure phase and the transition pressure from  the $\gamma$ and $\beta$ phases to $\alpha$ is determined to be about 1 GPa.
The electronic band structures 
of each phase at their own equilibrium volume are determined using 
the quasiparticle self-consistent (QS) $GW$ method as well as using the $0.8\Sigma$ approach in which the QS$GW$ self-energy is reduced by a factor 0.8 to correct for the underscreening of $W$ in QS$GW$. The 
effective masses of the band edges and the nature of the band gaps 
are presented. The lowest energy $\gamma$ phase is found to have a pseudodirect gap of 7.69 eV. The gap is direct at $\Gamma$ but corresponds to a dipole forbidden transition. The imaginary part of the dielectric function and the absorption coefficient are calculated in the long-wavelength limit and the random phase approximation, without local field or electron-hole interaction effects  for each phase and their anisotropies are discussed. Si doping on the Al site 
is investigated as a possible $n$-type dopant in $\gamma$-LiAlO$_2$ using a 128 atom supercell corresponding to 3.125 \% Si on the Al sublattice in the generalized gradient approximation and a smaller 16 atom cell with 25 \% Si in the QS$GW$ approximation.  The Si is found to significantly perturb the conduction band  and lower the gap but a clearly separated deep donor defect level is not found. However, the donor binding energy is still expected to be relatively deep, of order a few 0.1 eV in the hydrogenic effective mass approximation. 
\end{abstract}
\maketitle


\section{\label{sec:level1} Introduction}
LiAlO$_2$ is a ceramic material that is known to occur in at least five crystal structures: rhombohedral $\alpha$-LiAlO$_2$ (space group R$\overline{3}$m, $\#166$) \cite{Ma}, orthorhombic $\beta$-LiAlO$_2$ (space group $Pna2_1$, $\#33$) \cite{Vanfleet}, tetragonal $\gamma$-LiAlO$_2$ (space group $P4_12_12$ (\#92)) \cite{Marezio}, and tetragonal $\delta$-LiAlO$_2$ (space group $I4_1/amd$, $\#141$)  \cite{Li, Lei}. Among these, both $\beta$ and $\gamma$ forms are tetrahedrally coordinated while $\alpha$ and $\delta$ are octahedrally coordinated. The $\delta$ form is essentially a slightly disordered rocksalt type phase, in which the 4a Wyckoff positions are about 80 \% occupied with Al, 20 \% Li and with the roles of Al and Li 
reversed on the 4b positions. A fully disordered $\varepsilon$ cubic phase has also been reported.\cite{Lei}  A 48 atom cell 
with spacegroup  $P\overline{4}m2$ is listed in Materials Project and provides an approximate computational model for these disordered rocksalt type phases.

\begin{figure*}
    \includegraphics[width=18cm]{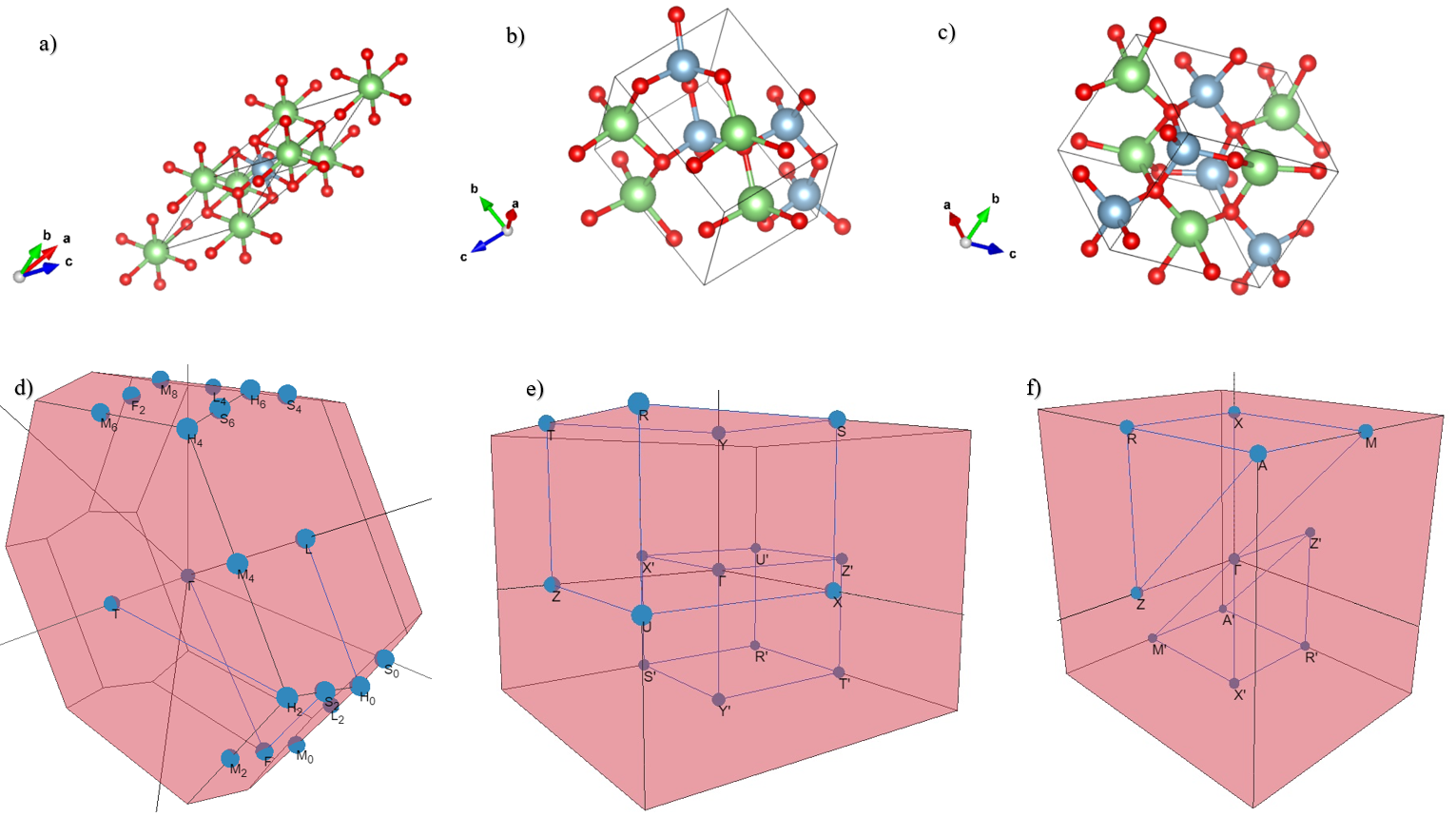}
    \caption{Crystal structures (a-c) (made with VESTA \cite{VESTA}) and Brillouin zones (d-f) (made with SeeK-path\cite{seekpath}) of  $\alpha$, $\beta$ and $\gamma$ LiAlO$_2$. Spheres indicate
      Li (green) Al (blue), O (red).\label{strucbz}}
\end{figure*}

LiAlO$_2$ has been investigated for applications in tritium breeding in fusion reactors \cite{Charpin, Khomane}, as a matrix in molten carbonate fuel cells \cite{Batra}, and as a substrate for GaN light-emitting diodes (LED) due to the small lattice mismatch between $\gamma$-LiAlO$_2$ and GaN \cite{Ke}. 
Moreover, LiAlO$_2$ is closely related to LiGaO$_2$---another ceramic material that is also known to adopt $R\overline{3}m$ and $Pna2_1$ structures and that has recently become of interest as a possible ultra-wide band gap (UWBG) semiconductor due to suggestions that LiGaO$_2$ could be $n$-type doped by silicon or germanium \cite{Boonchun1, Boonchun2, Lenyk, Skachkov, Dabsamut2020}. Various experimental works have found the band gap of LiGaO$_2$ in the range of $5.26$-$5.6$ eV \cite{Ohkubo, Wolan, Chen, Johnson}, and quasiparticle self-consistent $GW$ (QS$GW$) \cite{Kotani} calculations predict a gap of $5.81$ eV \cite{Radha}. We expect LiAlO$_2$ to have an even larger band gap than LiGaO$_2$.

LiAlO$_2$ has recently also been considered
in the context of Li-ion batteries. It was used as a coating to protect LiCoO$_2$ \cite{Cao2005} but
Li was also found to diffuse in LiAlO$_2$ itself at high temperatures\cite{Wiedemann2016,Wiedemann2016a}
and solid solutions of LiAlO$_2$ with
LiMO$_2$, where  M is a transition metal, were proposed by Ceder \etal\cite{Ceder1998}. LiAlO$_2$
has also been used as an additive in composite electrolytes \cite{Lakshman2004}. For a more complete
literature overview of these recent applications, see Singh \etal\cite{Singh2018}.
The phase transitions
of LiAlO$_2$ at high temperature are important in this context and were studied by Singh \etal\cite{Singh2018}.
as well as the pressure induced transitions we consider here. 

The electronic structure and phase transitions of LiAlO$_2$ have already been studied to some extent.
The structure of $\gamma$-LiAlO$_2$ at ambient
pressure and temperature was determined experimentally by Marezio via Cu $K\alpha$ and Mo $K\alpha$ radiation photographs \cite{Marezio}. The $\gamma \rightarrow \delta$ phase transition has been studied experimentally \cite{Li, Lei}, and Sailuam \textit{et al.} \cite{Sailuam}  have carried out a first-principles computational study of the band structures and pressure-induced $\gamma \rightarrow \delta$ phase transition \cite{Sailuam}. Ma \textit{et al}. have done a first-principles study of $\alpha$-LiAlO$_2$ \cite{Ma}. Singh \etal \cite{Singh2018} did an extensive computational study of the temperature-pressure phase diagram and also reported electronic band structures of the various phases.
However, there still remain open questions, especially on the electronic structure.
With the exception of the study by Sailuam\cite{Sailuam}, prior first-principles works have largely used density functional theory (DFT) methods with local or semilocal exchange correlation functionals, which are well-known to significantly underestimate band gaps. Sailuam \textit{et al.}\cite{Sailuam} obtained a gap of 6.56 eV for $\gamma$-LiAlO$_2$ using
the Heyd-Scuseria-Ernzerhof (HSE) hybrid functional, much larger than previous DFT studies. 
We seek to further improve upon these estimates using the more accurate QS$GW$ method.
Not only the band gap but other details of the band structure, such as the direct or indirect nature of the gap and the effective masses are relevant to the potential semiconductor applications, which have not yet
received attention. Thus, our paper is focused on the electronic structure and its properties relevant to
potential but not yet explored applications of the material as an active semiconductor.  Therefore
we will also briefly consider the possibility of doping. 

In relation to the band gap and opto-electronic  properties, we point that that  
Huang \textit{et al.} \cite{Huang2008} determined an optical absorption onset from transmission of VIS-UV to occur at 191 nm indicating  a gap of $\sim$6.5 eV consistent with  Zou \textit{et al.}\cite{Zou2005} who also measured a drop in transmission near 190 nm.  More recently, Holston \textit{et al.}\cite{Holston15,Holston2015} measured optical defects in LiAlO$_2$
induced by radiation of doping with Cu. They also do not see any absorption in as-grown crystals above 200 nm wavelength. From these 
we can deduce that the optical gap appears to be near 6.5 eV 
at room temperature. Besides optical properties, they also report electron paramagnetic resonance (EPR) spectra of these defects.

\par
The lowest-energy and most thermally stable form of LiAlO$_2$ is the $\gamma$ phase, but coexistence of the $\alpha$ and $\gamma$ phases is a common result of syntheses performed below $973$ K \cite{Becerril}, and surface contamination of bulk $\gamma$-LiAlO$_2$ with $\beta$-LiAlO$_2$ is known to occur as a result of matching between the $a$ and $b$ axes of the orthorhombic $\beta$ phase with the $a$ and $c$ axes of the tetragonal $\gamma$ phase \cite{Vanfleet}. Therefore, we consider each of the $\alpha$, $\beta$, and $\gamma$ phases. Additionally, the $R\overline{3}m$ structure was found to be a high-pressure phase of LiGaO$_2$ \cite{Radha}, so it is also interesting to study $\alpha$-LiAlO$_2$ for the sake of comparing these two closely related materials. 
We omit here the $\delta$-phase because it is actually a disordered phase, which requires 
larger cells to model, and the octahedral coordination is already represented by the $\alpha$-phase.
While other work in literature \cite{Singh2018} has already studied phase transitions extensively,
it is useful to compare our results to their work.
The structures studied in this paper and their Brillouin zones are shown
in Fig. \ref{strucbz}.

\section{Computational Methods}
The structure relaxation calculations in this work are performed using the Quantum Espresso pseudopotential plane-wave implementation of density functional theory (DFT)  \cite{Giannozzi}. We use projector augmented-wave pseudopotentials for all Quantum Espresso calculations \cite{Blochl}. 
The exchange-correlation  energy was treated in the generalized gradient approximation (GGA) using the Perdew-Burke-Ernzerhof (PBE) parametrization\cite{PBE}. For each of the $\alpha$, $\beta$, and $\gamma$ phases, we start with the lattice parameters provided by Materials Project \cite{MP} where available.  The unit cell is relaxed at a range of volumes around the equilibrium volume, and the cohesive energy per formula unit is calculated from the total energy at self-consistency and the reference energies of the free atoms (not including spin-polarization corrections of the atom).  Bulk moduli and first derivatives of the bulk moduli w.r.t. pressure are then extracted for each phase by fitting the corresponding energy vs. volume points to the energy curve obtained from integrating the Birch-Murnaghan equation of state \cite{Birch} via nonlinear least squares. Transition pressures between different phases are then extracted from these fits via the common tangent method. 
The structural relaxation results for the internal coordinates were further checked with the all-electron full-potential linearized muffin-tin orbital method (FP-LMTO) as implemented in the {\sc Questaal} code\cite{Pashov}  while 
keeping the lattice constant ratios fixed.  The obtained energy ordering of the phases was found to be consistent with {\sc Quantum Espresso} results 
provided the same muffin-tin radii were kept for all phases. 
\par
To overcome the limitations of semilocal DFT for band gaps, 
we use the many-body-perturbation theoretical $GW$ method.\cite{Hedin65}
The band structure calculations performed  here use  the quasiparticle self-consistent version of the $GW$ method\cite{Kotani} (QS$GW$)  as implemented in the {\sc Questaal} code.\cite{Pashov} The $G$ in $GW$ represents the one-particle Green's function, and the $W$ represents the screened Coulomb interaction. The QS$GW$ approach is independent of the DFT starting point Hamiltonian $H^0$, because a non-local exchange-correlation
potential is extracted from the $GW$ self-energy,
 in the form of its matrix in the basis set of $H^0$ eigenstates  $\tilde\Sigma_{nm}=\frac{1}{2}\mathrm{Re}\left[ \Sigma_{nm}(\epsilon_n)+\Sigma_{nm}(\epsilon_m)\right]$  where
 Re indicates taking the hermitian part, 
and is used to update $H^0$ and iterated to convergence. At each iteration $\Sigma(\omega)=iG^0(\omega)\otimes W(\omega)$ with $\otimes$ standing for convolution. At convergence, the quasiparticle energies of $GW$ become 
equal to the Kohn-Sham eigenvalues of the updated $H^0$. 
This approach is known to give much more accurate band gaps  than 
semilocal DFT but is known to systematically overestimate the band gaps 
slightly because the screening of $W$ is calculated in the random 
phase approximation and thereby underestimates screening by not including
electron-hole interaction effects.  This shortcoming 
can be overcome by including ladder diagrams\cite{Cunningham18,Cunningham21} but this is still a rather expensive approach and it was found  that reducing the self-energy correction $\tilde\Sigma-v_{xc}^{DFT}$ by a universal factor 0.8, which we call $0.8\Sigma$ provides an practical alternative \cite{Deguchi16,Bhandari18}.

The main factors that determine the accuracy and 
convergence of the FP-LMTO implementation of the QS$GW$ method\cite{Kotani} include the basis-set choice and the {\bf k}-point mesh on which the self-energy is calculated. In the FP-LMTO method we use augmented spherical waves with smoothed Hankel functions as radial functions outside the spheres  and typically use a basis set with two sets of Hankel function energies and smoothing radii. Here we use an $spdfspd$ basis set on each of the atoms, which 
means one set of smoothed Hankel functions includes orbitals 
up to $f$ and the second up to $d$ spherical harmonics. The smoothing radii 
and energies are chosen in a standard way by fitting these 
radial wave functions to the tail of the free atom eigenstates. 
 The basis functions are expanded to $l_{max}=4$ within each augmentation sphere. To further test the convergence, we added 
high-energy $s$ and $p$ local orbitals but found them to 
affect the band gap negligibly.
The two-point quantities, such as the bare and screened Coulomb interaction $W$ are represented in a separate basis set of Bloch functions which include products of partial waves inside the spheres and plane waves projected on the interstitial region. This basis set is far more efficient than a plane wave basis set at representing the screening within the 
Hilbert space of the bands of interest 
and thereby also reduces the need to include
high-lying empty states in calculating the polarization propagator 
$P$ used in $W=v+vPW$ with $v$ the bare Coulomb interaction and in 
the calculation of $\Sigma$. The mixed product basis set and interstitial plane wave 
basis functions in the QS$GW$ scheme are expanded 
to $G_{max}$ of 2.7 Ry$^{1/2}$ and 3.3 Ry$^{1/2}$, which 
are standard values. 

The final (i.e. after checking convergence) $\mathbf{k}$-meshes and energy cutoffs for $\Sigma$ used in the QS$GW$ calculations for each phase are given in Table \ref{Tab1}.   In the QS$GW$ iterations 
the $\tilde\Sigma$ is replaced by an average value between 
$E^{max}_{\Sigma}$ and $E^{max}_{\Sigma}-0.5$ Ry. The  atom-centered  LMTO basis set 
provides a natural interpolation scheme for the self-energy matrix
to a finer {\bf k}-point mesh used in the charge-potential self-consistency 
iterations and for obtaining the bands along the symmetry lines. 
It thus provides accurate band dispersions and also effective masses.  For details about the QS$GW$ implementation we refer the reader to the method description papers \cite{Kotani,Pashov}.

\begin{table}
\caption{Convergence parameters for QS$GW$ band structure calculations, $E^{max}_{\Sigma}$ is the maximum energy  up to which the self-energy $\Sigma(\omega)$ is calculated. \label{Tab1}}
\begin{ruledtabular}
\begin{tabular}{lll}
\textrm{Phase}&
$\mathbf{k}$\textrm{-Mesh}&
\textrm{$E^{max}_{\Sigma}$ (Ry)}\\
\colrule
$\alpha$ & 6 $\times$ 6 $\times$ 6 & 3.9\\
$\beta$ & 3 $\times$ 3 $\times$ 3 & 3.5\\
$\gamma$ & 3 $\times$ 3 $\times$ 3 & 3.4
\end{tabular}
\end{ruledtabular}
\end{table}
\section{Results}
\subsection{Equations of state and transition pressures}
We begin with an investigation of the structural properties and high-pressure phase transitions of $\alpha$, $\beta$, and $\gamma$-LiAlO$_2$. For the $\alpha$ and $\gamma$ phases, we use as initial input the lattice constants and site positions provided by Materials Project \cite{MP}. However, at the time of our investigation, Materials Project did not have structural information for $\beta$-LiAlO$_2$, so we instead use the data for LiGaO$_2$ in the $Pna2_1$ structure and then relax the unit cell substituting aluminum for gallium.

\begin{figure}
    \centering
    \includegraphics[width=0.45\textwidth]{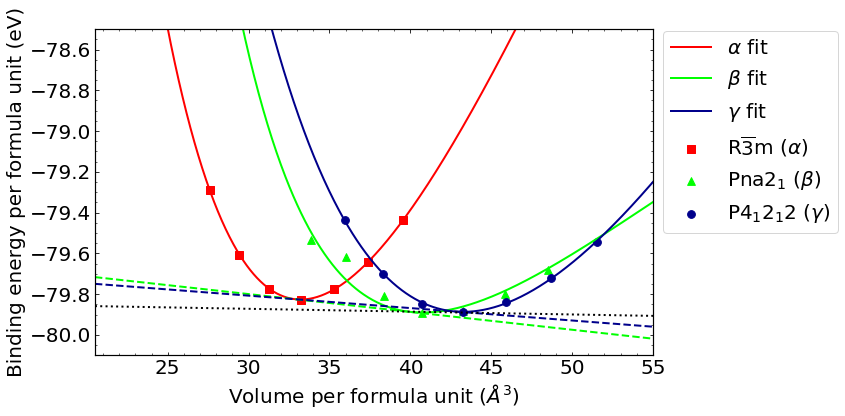}
    \caption{Directly calculated energy vs. volume points and equation of state fits for $\alpha$, $\beta$, and $\gamma$-LiAlO$_2$, along with common tangents corresponding to the $\beta \rightarrow \alpha$ (green dashed) and $\gamma \rightarrow \alpha$ (blue dashed ) and $\gamma\rightarrow\beta$ (black dotted) high-pressure phase transitions.}
    \label{fig:Fig1}
\end{figure}

\begin{table}
  \caption{Fitting parameters corresponding to Fig. \ref{fig:Fig1}, $B_0$, bulk moldules at equilibrium, $B_0^\prime$, pressure derivative of the bulk modulus at equilibrium volume, $p_t$ is the transition pressure.\label{Tab2}}
\begin{ruledtabular}
\begin{tabular}{lccc}
\textrm{Phase}&
$B_0$ (GPa)&
$B_0^\prime$&
$p_t$ (GPa)\\
\colrule
$\alpha$ & 136 & 3.64 & \\
$\beta$ & 70.5 & 6.01 & 1.4\footnote{$\beta\rightarrow\alpha$}\\
$\gamma$ & 90.2 & 3.14 & 0.98\footnote{$\gamma\rightarrow\alpha$}, 0.22\footnote{$\gamma\rightarrow\beta$}\\
\end{tabular}
\end{ruledtabular}
\end{table}
 
First, we discuss the fitted energy vs. volume curves shown in Fig. \ref{fig:Fig1} and transition pressures between structures. We fit our directly calculated energies as functions of volume to the energy obtained from integrating the Birch-Murnaghan equation of state, which is given by \cite{Birch}
\begin{eqnarray}\label{eq:Eq1}
    E(V) &= E_0 + \frac{9 V_0 B_0}{16} \bigg\{ \left[ \left( \frac{V_0}{V} \right)^{2/3} - 1 \right]^3 B_0^{'} \nonumber \\ 
    &+ \left[ \left( \frac{V_0}{V} \right)^{2/3} - 1 \right]^2 \left[6 - 4 \left( \frac{V_0}{V} \right)^{2/3} \right] \bigg\}
\end{eqnarray}
where $E$ is the total energy in the crystalline state, $E_0$ is the total energy of the neutral free atoms at rest, $V$ is the volume of the unit cell, $V_0$ is the equilibrium volume of the unit cell, $B_0$ is the bulk modulus, and $B_0'$ is the first derivative of the bulk modulus w.r.t. pressure.

We calculate the transition pressures between different structures using the common tangent construction, wherein the negative of the slope of the tangent line is the pressure required to achieve enthalpic equality between two crystal structures. A plot of the directly calculated points and equation of state fits is shown in Fig. \ref{fig:Fig1}, and values of the fitting parameters $B_0$ and $B_0'$ along with the transition pressures are given in Table \ref{Tab2}. From Fig. \ref{fig:Fig1}, it is evident that the $\alpha$ phase is a high-pressure form of the material. This result is analogous to that obtained by Radha \textit{et al}. from their analysis of LiGaO$_2$, which found the $R\overline{3}m$ structure to be a high-pressure phase of that material \cite{Radha}. Further, the $\gamma$ phase is the lowest-energy structure, though the difference between the energy minima in the $\beta$ and $\gamma$ phases is only $\sim10^{-4}$ eV, which is consistent with experimental observation of these two phases coexisting. Their lattice volume per formula unit are also close but slightly larger for $\gamma$, indicating that also $\beta$ could be stabilized under pressure.
The common tangents shown in Fig. \ref{fig:Fig1} indicate a tetrahedral $\rightarrow$ octahedral transition around 0.98 GPa for $\gamma\rightarrow\alpha$ and 1.4 GPa for $\beta\rightarrow\alpha$.

These results agree well with those of Singh \etal \cite{Singh2018} who also found the $\beta$
and $\gamma$ phases to have very close energy minima  and establish a $\gamma\rightarrow\alpha$ transition at $\sim$1.3 GPA. They obtain this by calculating the energy and enthalpy directly as function of pressure,
by explicitly imposing the pressure as a stress tensor as independent variable in their calculation,
whereas we start from the energy-volume curves and use the common tangent construction, but in principle, these
procedures should give equivalent results.  They also mention
that in the range $0-1.2$ GPa the $\beta$-structure already has lower enthalpy. We calculate a transition pressure of 0.22 GPa for $\gamma\rightarrow\beta$ based on the fitted equations of state. 

The study by Singh \etal \cite{Singh2018} is more complete than
ours in terms of the phase diagrams by their inclusion of temperature and entropy effects
and by also considering the higher pressure transitions to the $\delta$-phase but for low
temperatures our results agrees well with  theirs.  Previously, Sailuam \cite{Sailuam}  studied the
$\gamma\rightarrow\delta$ transition and found a transition pressure of about 2-3 GPa depending on
which functional was used. This agrees qualitatively with Singh \etal.\cite{Singh2018}.
They did not consider the $\alpha$-phase but did evaluate energy barriers between the phases
as funtion of pressure. We preferred here to focus on the
basic tetrahedral to octahedral phase transition by considering the simpler $\alpha$ phase. After all,
the $\delta$-phase which is a disordered cubic phase with tetragonal distortion is not as well
established structurally or can only approximately be described by a small unit cell because of the
fractional occupations of lattice sites. 

For reference, we give the equilibrium lattice constants of each phase 
in Table \ref{tab:latconst} as obtained from the Quantum Espresso 
minimization. These are used later for the electronic structure on which we focus. For the $\alpha$-phase,
we here use the primitive rhombohedral cell parameters $a$ and the opening angle between the three
equal size lattice vectors. To compare with the hexagonal conventional cell, used in the work by
Singh \etal\cite{Singh2018}, which is three times larger in volume, note that  $a_h=\sqrt{2}a_r\sqrt{1-\cos{\alpha}}$ $c_h=3a_r\sqrt{1+2\cos{\alpha}}/\sqrt{3}$.
This gives $a_h=2.82$\AA\ $c_h=14.36$ \AA, in good agreement with the values by Singh \etal.\cite{Singh2018}.

\begin{figure*}
\centering
\includegraphics[width=\textwidth]{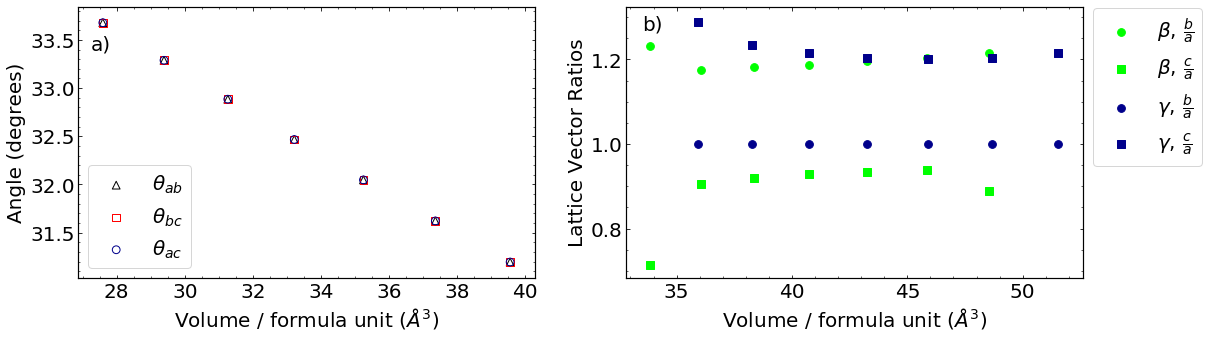}
\caption{\label{fig:Fig2} a) Lattice angles vs. volume for $\alpha$-LiAlO$_2$ (we use $\theta_{ab}$, $\theta_{bc}$, and $\theta_{ac}$ to denote the lattice angles instead of the traditional $\alpha$, $\beta$, and $\gamma$ to avoid confusion with the phase labeling of LiAlO$_2$). b) $\frac{b}{a}$ and $\frac{c}{a}$ ratios vs. volume for tetrahedrally coordinated $\beta$ and $\gamma$-LiAlO$_2$.}
\end{figure*}

\begin{table*}
\caption{Space groups, lattice constants, volume per  formula uniy, and Wyckoff positions for different phases of LiAlO$_2$.}
    \centering
    \begin{ruledtabular}
    \begin{tabular}{l|ccc}
   structure  &   $\alpha$ & $\beta$  & $\gamma$\\ 
   space group          &  $R\bar{3}m$ & $Pna2_1$ & $P4_12_12$ \\ \hline
                        & $a=5.06$ {\AA}         & $a=5.29$ {\AA}         & $a=5.24$ {\AA}   \\
   & $\alpha=32.47^\circ$ & $b=6.28$ {\AA} &  $c=6.31$ {\AA}\\
                      &            & $c=4.90$ ${\AA}$ & \\ 
   &         $V=33.2$ \AA$^3$            &  $V=40.7$ \AA$^3$                 &$V=43.3$ \AA$^3$ \\ \hline
  
                    Li &    $1a$     &   $4a$    &   $4a$   \\
                       & $(0,0,0)$  &   $(x=0.0837,y=-0.3761,z=-0.0033)$    &  $(x=-0.1858,y=-0.1858,z=0)$      \\   
                    Al &  $1b$ &  $4a$ & $4a$\\
                       & ($\frac{1}{2},\frac{1}{2},\frac{1}{2})$ & $(x=0.0786,y=0.1262,z=0.0048)$& $(x=0.1768,y=0.1768,z=0)$\\
                     O & $2c$ & $4a$ & $8b$\\
                       & $(\pm u,\pm u, \pm u)$ & $(x=0.0614,y=0.1053,z=0.3654)$ & $(x=0.3392,y=0.2904,z=-0.2271)$\\
                       & $u=0.2381$  & & \\
                     O$_{II}$ & &  $4a$ &  \\
                              & & $(x=0.1003,y=-0.3528,z=0.4062)$ & \\ 
    \end{tabular}
    \end{ruledtabular}
    \label{tab:latconst}
\end{table*}

It is also interesting to study how the lattice parameters themselves change as functions of unit cell volume. 
For the rhombohedral $\alpha$-structure the volume of the unit cell is given by $V=a^3\sqrt{1-3 \cos^2{\alpha}+2\cos^3{\alpha}}$. 
Part (a) of  Fig. \ref{fig:Fig2} shows that the opening angle $\alpha$ between each pair of lattice vectors, decreases with increasing volume. For the $\beta$-structure we can see in panel (b)
that $b/a$ and $c/a$ stay more or less constant as the volume 
is decreased until the volume approaches the region where the
phase transition to the $\alpha$-phase occurs. The sudden 
change in these ratios indicates the incipient instability of the 
$\beta$-structure and is possibly related to the transition path 
between wurtzite and rocksalt suggested in Ref. \onlinecite{Limpijumnong2001}. Likewise in the $\gamma$-structure, 
the $c/a$ ratio is seen to increase when approaching the transition 
volume but otherwise stays constant and $b/a$ stays equal to 1 
as required by symmetry.  We have checked that during the 
relaxation procedure, the symmetries required by each lattice 
were maintained.

\begin{figure}[t]
    \centering
    \includegraphics[width=0.45\textwidth]{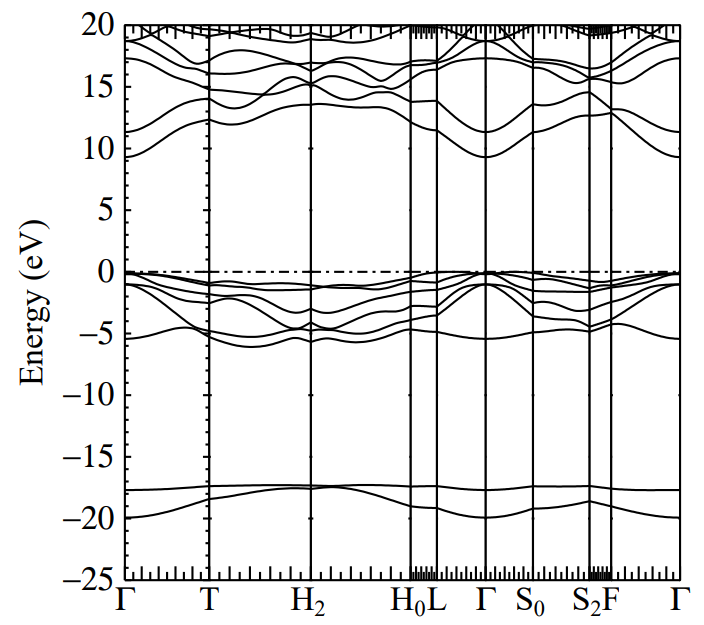}
    \caption{QS$GW$ $0.8 \Sigma$ band structure of $\alpha$-LiAlO$_2$ spanning a wide range of valence and conduction states.}
    \label{fig:Fig3}
\end{figure}

\begin{figure}[h]
    \centering
    \includegraphics[width=0.45\textwidth]{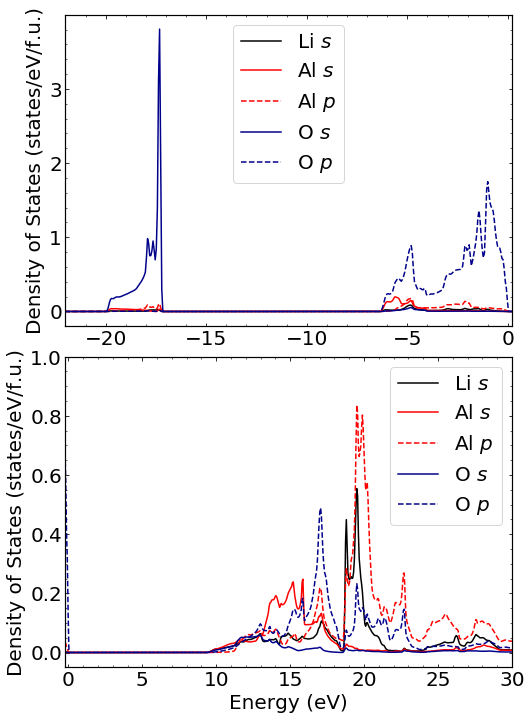}
    \caption{$\ell$-resolved partial densities of states in $\alpha$-LiAlO$_2$ in the (top) valence and (bottom) conduction bands.}
    \label{fig:Fig4}
\end{figure}

\subsection{Band structures}
For our final band structure calculations, we use the lattice parameters corresponding to the lowest-energy point on each of the equation of state fits of Fig. \ref{fig:Fig2}, i.e. those given in Table \ref{tab:latconst}.
\par
The QS$GW$ $0.8 \Sigma$ approximation band structure of $\alpha$-LiAlO$_2$ is shown over a wide energy range in Fig. \ref{fig:Fig3}, with corresponding plots of the partial densities of states (resolved by $\ell$) in both the valence and conduction bands shown in Fig. \ref{fig:Fig4}. We find that the $\alpha$ phase has an indirect gap of $9.30$ eV, with the conduction band minimum occurring at the $\Gamma$-point. The indirect nature is already present also in the GGA-band structure. From the partial densities of states, we can discern that the deep-lying bands spanning $\sim -20$ to $\sim -17.5$ eV are those derived primarily from the oxygen-$2s$ orbitals. The higher-lying valence states from $\sim -5$ to $0$ eV are primarily due to the oxygen-$2p$-derived bands, though there are contributions from both aluminum-$3s$ and $3p$ as well because these are bonding states with the cation atomic orbitals.  
The lowest-lying (around $10$ eV) conduction states consist of a mixture of lithium-$2s$, (antibonding) oxygen-$2s$ and $2p$, and aluminum-$3s$ orbitals. At higher energy, there are peaks corresponding to significant lithium-$2s$, oxygen-$2p$, and aluminum-$3p$ contributions. This confirms the ionic picture in which Li donates its electrons to the oxygen. 
\par
\begin{figure}[t]
    \centering
    \includegraphics[width=0.45\textwidth]{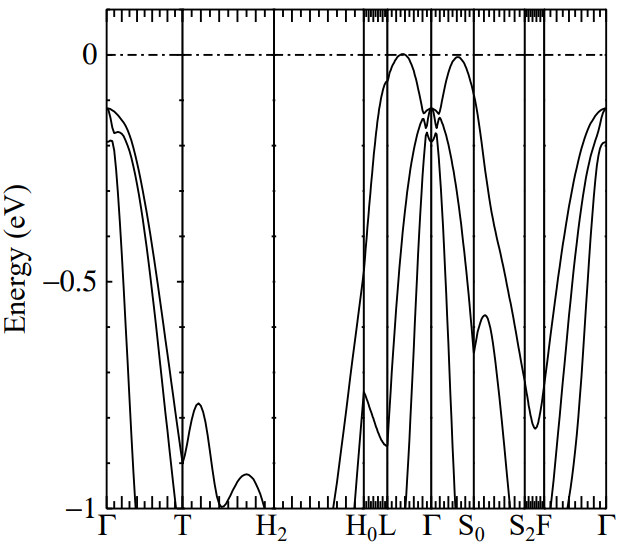}
    \caption{Magnified view of the high-lying valence states of the QS$GW$ $0.8 \Sigma$ band structure of $\alpha$-LiAlO$_2$.}
    \label{fig:Fig5}
\end{figure}

We now turn to a magnified view of the valence states near the Fermi level, shown in Fig. \ref{fig:Fig5}. 
This show that two almost equal energy valence band maxima (VBM) occur between $\Gamma$ and L and $\Gamma$ and $S_0$. The nomenclature for the high-symmetry points follows the convention of the 
Bilbao crystallographic server  website  (\url{https://www.cryst.ehu.es/}) and is also given
in Fig. \ref{strucbz}. The conduction band minimum (CBM) 
meanwhile is at $\Gamma$ and the material thus has an indirect 
band gap. At $\Gamma$ the  VBM is doubly degenerate and 
has $E_u$ symmetry of the $D_{3d}$ point group, which is $(x,y)$-like while the state below it is non-degenerate with $A_{1u}$ symmetry, which is $z$-like. This indicates that direct
vertical transitions from the VBM at $\Gamma$ to the conduction band which has $A_{1g}$ symmetry ($s$-like) are dipole-allowed for polarization perpendicular to the three-fold symmetry axis, while the transitions from the crystal field split-off state will occur for polarization along the symmetry axis. This is confirmed by the optical absorption calculations discussed below. However, the indirect gap 
is about 0.1 eV lower than the direct gap. One may notice some avoided
band crossings just below the VBM along $\Gamma-T$, $\Gamma-L$ and $\Gamma-S_0$.
These were checked by using a fine spacing of the {\bf k}-points along the lines
and indicate that these bands belong to the same irreducible representation
of the group of {\bf k} along these lines and
can therefore not cross. 

\begin{figure}[h]
    \centering
    \includegraphics[width=0.45\textwidth]{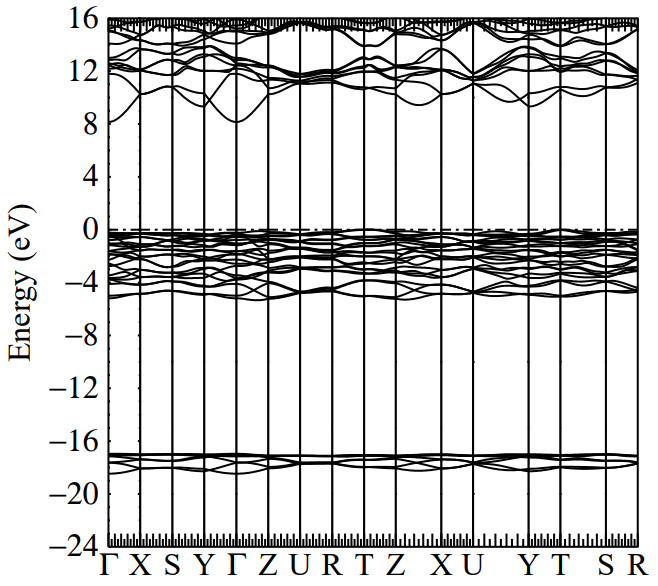}
    \caption{QS$GW$ $0.8 \Sigma$ band structure of $\beta$-LiAlO$_2$ spanning a wide range of valence and conduction states.}
    \label{fig:bandsbeta}
\end{figure}

\begin{figure}[h]
    \centering
    \includegraphics[width=0.45\textwidth]{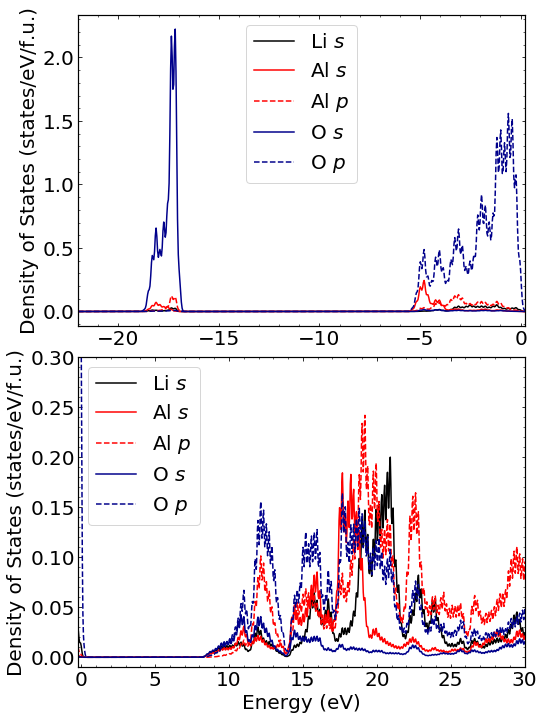}
    \caption{$\ell$-resolved partial densities of states in $\beta$-LiAlO$_2$ in the (top) valence and (bottom) conduction bands. A Gaussian broadening with a width of $0.1$ eV was applied to these spectra to better distinguish some of the features.}
    \label{fig:betaPDOS}
\end{figure}

\begin{figure}[h]
    \centering
    \includegraphics[width=0.45\textwidth]{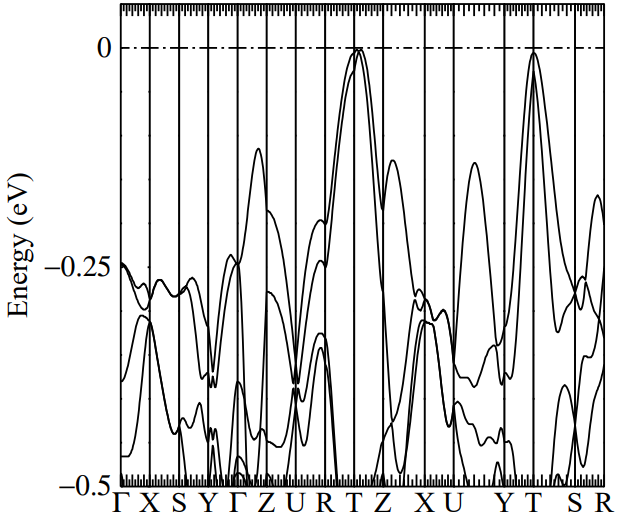}
    \caption{Magnified view of the high-lying valence states of the QS$GW$ $0.8 \Sigma$ band structure of $\beta$-LiAlO$_2$.}
    \label{fig:bandsbetazoom}
\end{figure}

\begin{figure}[h]
    \centering
    \includegraphics[width=0.45\textwidth]{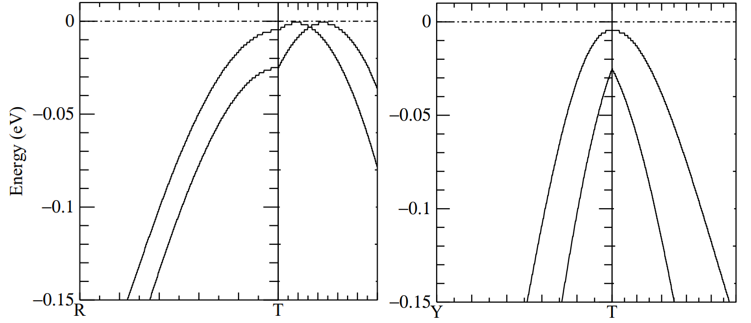}
    \caption{Close zoom-ins on the saddle point at $T$ and the actual VBM along $T - Z$ in the valence bands of $\beta$-LiAlO$_2$. The left panel shows the top-most bands along $R - T - Z$ and the right panel shows the top-most bands along $Y - T - S$.}
    \label{fig:bandsbetasaddle}
\end{figure}

The band structure of the $\beta$-structure is shown in Figs. 
\ref{fig:bandsbeta},\ref{fig:bandsbetazoom}, and \ref{fig:bandsbetasaddle}. The $\ell$-resolved partial densities of states are shown in Fig. \ref{fig:betaPDOS}.
The overall orbital character of the bands is the same as 
in the $\alpha$-structure. The zoom in near the VBM shows again 
an indirect band gap, and which is also already present in the GGA. 
The VBM occurs near the point $T$, 
which is  $(0,0.5,0.5)$ in  units of the reciprocal lattice vectors.
An even closer zoom shown in Fig. \ref{fig:bandsbetasaddle} right near the $T$ point in the directions $T-Z$
and $T-R$ shows that $T$ is a saddle point with a minimum in the $T-Z$
direction and a maximum along  $T-R$. The line along $T - Z$ contains two maxima close in energy to one another. It also is at a maximum in the $T-Y$ direction. Thus, great care is required to determine the effective mass tensor at the actual VBM. 
The VBM at $\Gamma$ lies $\sim 0.24$ eV below it.
The valence bands at $\Gamma$ from the highest one and down 
have irreducible symmetries, $a_2$, $a_1$, $b_1$, $b_2$ which 
are respectively forbidden, allowed for $z$, $x$, $y$ polarization
for transitions the the $a_1$ symmetry CBM at $\Gamma$. These symmetry labels were determined by inspection of the eigenvectors.

\begin{figure}[h]
    \centering
    \includegraphics[width=0.45\textwidth]{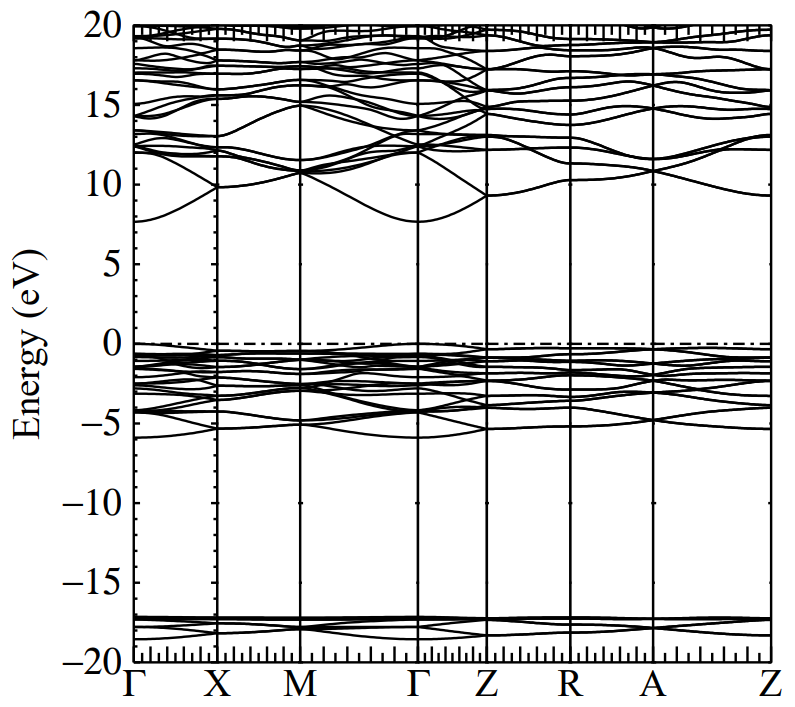}
    \caption{QS$GW$ $0.8 \Sigma$ band structure of $\gamma$-LiAlO$_2$ spanning a wide range of valence and conduction states.}
    \label{fig:bandsgamma}
\end{figure}

\begin{figure}[h]
    \centering
    \includegraphics[width=0.45\textwidth]{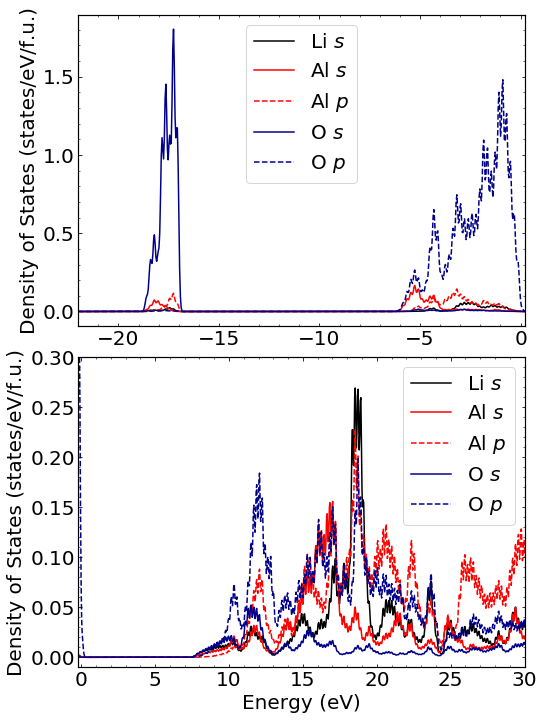}
    \caption{$\ell$-resolved partial densities of states in $\gamma$-LiAlO$_2$ in the (top) valence and (bottom) conduction bands. A Gaussian broadening with a width of $0.1$ eV was applied to these spectra to better distinguish some of the features.}
    \label{fig:gammaPDOS}
\end{figure}

\begin{figure}[t]
    \centering
    \includegraphics[width=0.45\textwidth]{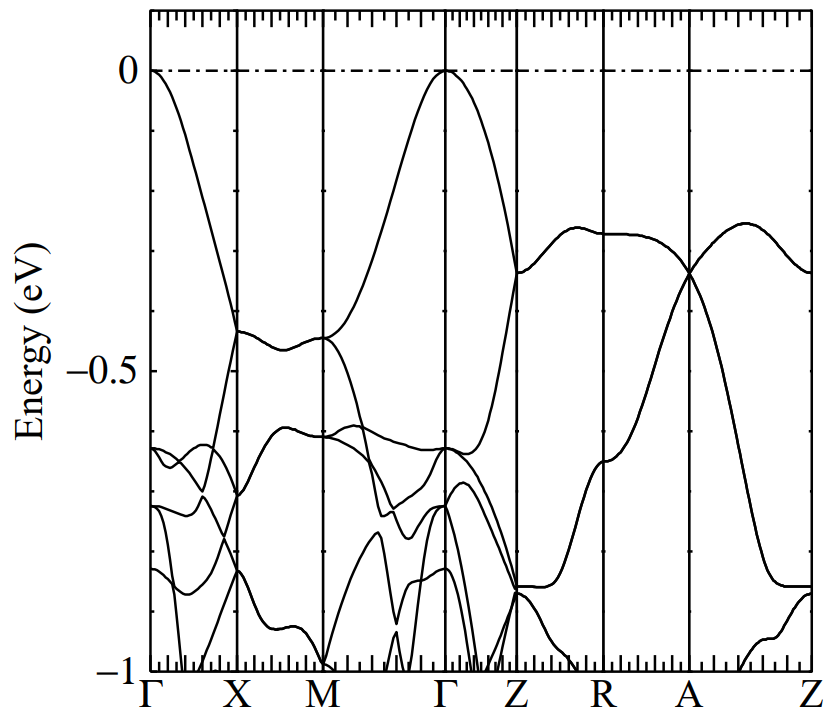}
    \caption{Magnified view of the high-lying valence states of the QS$GW$ $0.8 \Sigma$ band structure of $\gamma$-LiAlO$_2$.}
    \label{fig:bandsgammazoom}
\end{figure}

Finally for the $\gamma$ structure, the band structure is shown in 
Figs. \ref{fig:bandsgamma} and \ref{fig:bandsgammazoom}, with the $\ell$-resolved partial densities of states in Fig. \ref{fig:gammaPDOS}. The overall orbital character of the bands is similar to the $\alpha$ and $\beta$ phases. 
The zoom up show that this phase has a direct band gap. 
The CBM has $A_1$ symmetry but  the  VBM has $B_1$ symmetry in the 
point group $D_4$.
This implies that the gap is pseudodirect. In other words it is direct
but dipole forbidden. The next lower VBM at $\Gamma$ is double degenerate and therefore has $E$ symmetry but lies about 0.63 eV lower.   However, along $\Gamma-X$ the group of ${\bf k}$ is $C_2$
and contains the $C_{2x}$ symmetry axis under which $B_1$ is even. 
This means direct vertical transitions for states along the $\Gamma-X$ axis become allowed for $z$-polarization because both CBM and VBM 
at these points and $z$ are even or belong to the $A$ irreducible representation of $C_2$.  Meanwhile along $\Gamma-M$,
the group of ${\bf k}$ is also $C_2$ but contains the $C_2$ axis along the (110) direction, under which $B_1$ is odd. This implies that vertical transitions from the valence band along $\Gamma-M$ to the conduction 
band become allowed for $x$ or $y$ polarization because the CBM is still even under that $C_2$-operation. In other words, the VBM has 
irreducible representation $B$ while the CBM has irrep A and transitions are allowed for $x$ or $y$ because these belong to the $B$ irrep.   Along $\Gamma-Z$ the group of ${\bf k}$ is $C_4$ and the top VB belongs to the $B$ irreducible representation and thus the transitions are still forbidden.  We will see that thus both $x$ and 
$y$ polarization should be allowed for energies slightly above the direct gap at $\Gamma$ but also very close to it transitions for $z$ polarization. 

The band gaps of all three phases are summarized in Table \ref{tab:gaps}.
Our GGA  band gaps of 4.48 eV for $\gamma$ and 6.18 eV for $\alpha$ agree reasonably with Singh \etal's\cite{Singh2018} 4.7 eV
and 6.2 eV respectively.  
Qualitatively, the differences in gap between the different structures agree well with prior work
in the sense that the $\beta$ structure has only slightly higher gap than $\gamma$ but the octahedral
phases have a significantly larger gap.  The QS$GW$ gaps of course are significantly larger and are more
reliable. In terms of details of the band structure, prior work missed the indirect nature of the gap
in the $\alpha$ and $\beta$ structures. For the $\alpha$-structure, they used the conventional hexagonal cell
and hence show the bands in the corresponding hexagonal Brillouin zone. They may thus have missed the points
where the actual VBM occurs. Secondly, for all three phases, they  only show a band structure on large
energy scale, were the top valence bands are very flat and therefore it is difficult to ascertain where the actual VBM occurs.

\begin{table}
\caption{Band gaps of $\alpha$, $\beta$, and $\gamma$-LiAlO$_2$ from various approximations, of which QS$GW$ $0.8 \Sigma$ is expected to be the most accurate.\label{tab:gaps}}
\begin{ruledtabular}
\begin{tabular}{llll}
\textrm{Phase}&
GGA&
QS$GW$&
QS$GW$ $0.8 \Sigma$\\
\colrule
$\alpha$ & 6.18 eV & 10.1 eV & 9.30 eV\\
$\beta$ & 4.93 eV & 8.94 eV & 8.16 eV\\
$\gamma$ & 4.48 eV & 8.47 eV & 7.69 eV
\end{tabular}
\end{ruledtabular}
\end{table}

\subsection{Effective masses}
The curvature of the bands near the band edges provide the 
effective masses, which are important for transport properties. 
They are summarized in Table \ref{tab:effmass}.

\begin{table}[h]
\caption{Effective electron masses at the VBMs and CBMs of $\alpha$, $\beta$, and $\gamma$-LiAlO$_2$, extracted from the band curvatures near the extremal points.\label{tab:effmass}}
\begin{ruledtabular}
\begin{tabular}{llll}
\textrm{Phase/Extreme Point}&
$m_{xx}$ ($m_e$)&
$m_{yy}$ ($m_e$)&
$m_{zz}$ ($m_e$)\\
\colrule
$\alpha$ (VBM)\footnote{Principal values of the mass tensor. The masses labeled $x$ and $z$ actually correspond to principal axes 15 $^\circ$ from the crystal  axes.} & -6.6 & -0.69 & -2.8\\
$\alpha$ (CBM) & 0.47 & 0.47 & 0.58\\
$\beta$ (VBM) & -1.3 & -0.57 & -0.94\\
$\beta$ (CBM) & 0.43 & 0.39 & 0.36\\
$\gamma$ (VBM) & -1.7 & -1.7 & -2.9\\
$\gamma$ (CBM) & 0.41 & 0.41 & 0.44
\end{tabular}
\end{ruledtabular}
\end{table}

As expected from the symmetries of the $\alpha$ and $\gamma$ phases, the $\Gamma$-point CBM of the $\alpha$ phase and CBM and VBM of the $\gamma$ phase have effectively one transverse and one longitudinal effective mass. The valence band masses are significantly higher than the conduction band masses as is evident from the flat top valence band.  For the $\beta$-phase they were determined by zooming in very closely to the actual VBM along $T-Z$, as shown 
in Fig. \ref{fig:bandsbetasaddle}. Note that we here give electron masses, so the negative signs for the VBM indicate positive hole-masses.

\subsection{Optical absorption}
The imaginary part of the dielectric function $\varepsilon_2(\omega)$, which is proportional to the 
optical absorption coefficient was calculated in the long-wavelength independent particle approximation, ignoring local field and 
excitonic effects, according to 
\begin{eqnarray}
\varepsilon_2(\omega)&=&\frac{8\pi^2e^2}{\Omega \omega^2}\sum_v\sum_c\sum_{{\bf k}\in BZ}f_{v{\bf k}}(1-f_{c{\bf k}}) \nonumber \\
&&|\langle \psi_{{v\bf k}}|[H,{\bf r}]|\psi_{c{\bf k}}\rangle|^2\delta(\omega-\epsilon_{c{\bf k}}+\epsilon_{v{\bf k}}). \label{eq:AdlerWiser}
\end{eqnarray}
where the commutator $[H,{\bf r}]$ gives the band velocity 
and includes the contributions from the non-local self-energy, 
$\epsilon_{v{\bf k}}$, $\epsilon_{c{\bf k}}$ are the valence 
and conduction band states at ${\bf k}$ obtained in the 
QS$GW$ method. The $f_{n{\bf k}}$ are the band occupation 
numbers (Fermi functions at zero temperature) and are $1$ for $n=v$ and $0$ for $n=c$. $\Omega$ is he volume of the unit cell. 

\begin{figure}[h]
    \centering
    \includegraphics[width=0.45\textwidth]{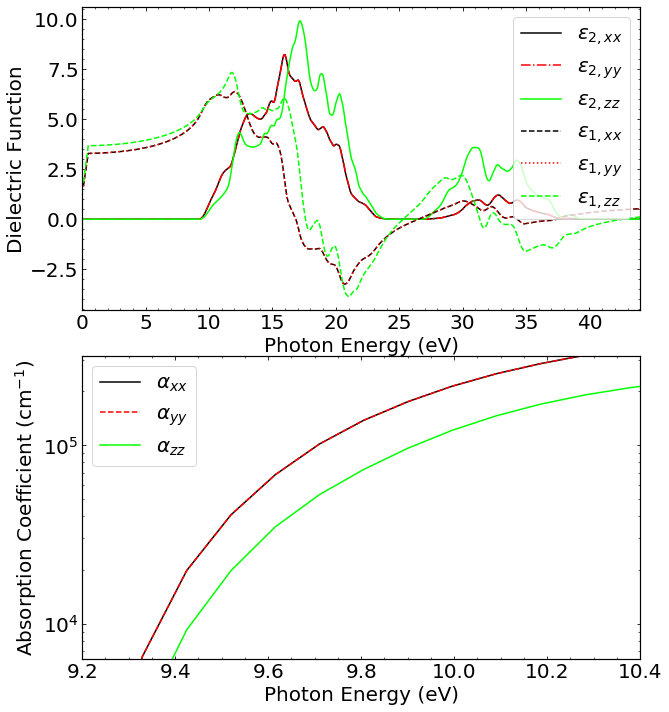}
    \caption{(Top) Real ($\epsilon_1$) and imaginary ($\epsilon_2$) components of the dielectric function for $\alpha$-LiAlO$_2$ and (bottom) optical absorption coefficients in logarithmic scale near the onset of absorption.}
    \label{fig:alphaDielectric}
\end{figure}

\begin{figure}[h]
    \centering
    \includegraphics[width=0.45\textwidth]{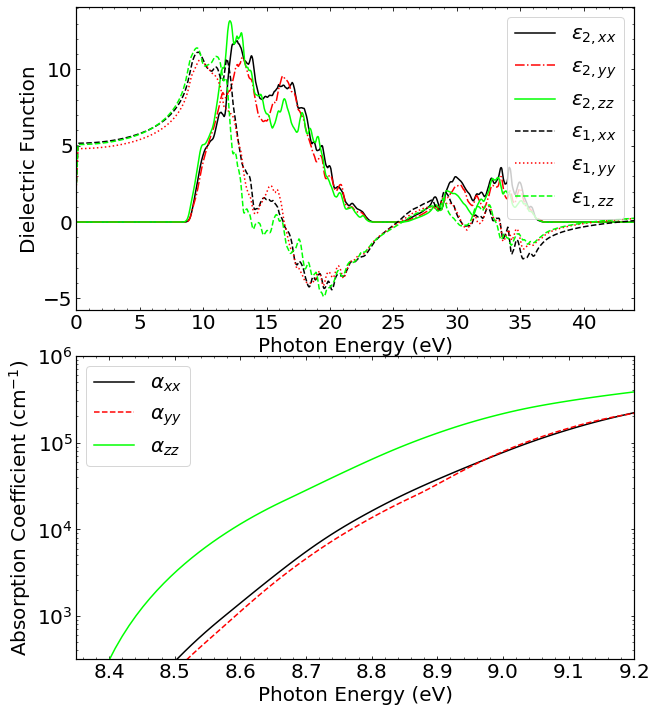}
    \caption{(Top) Real ($\epsilon_1$) and imaginary ($\epsilon_2$) components of the dielectric function for $\beta$-LiAlO$_2$ and (bottom) optical absorption coefficients in logarithmic scale near the onset of absorption.}
    \label{fig:betaDielectric}
\end{figure}

\begin{figure}[h]
    \centering
    \includegraphics[width=0.45\textwidth]{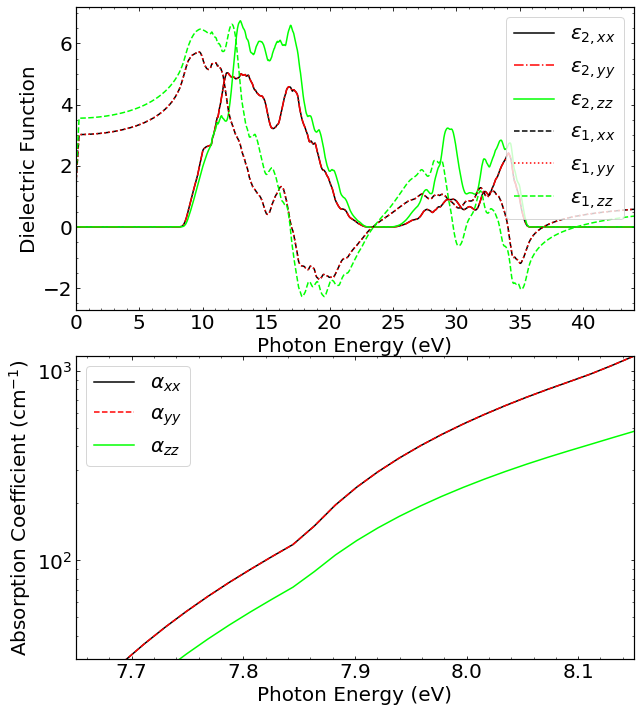}
    \caption{(Top) Real ($\epsilon_1$) and imaginary ($\epsilon_2$) components of the dielectric function for $\gamma$-LiAlO$_2$ and (bottom) optical absorption coefficients in logarithmic scale near the onset of absorption.}
    \label{fig:gammaDielectric}
\end{figure}

These functions are shown in Figs. \ref{fig:alphaDielectric}, \ref{fig:betaDielectric}, and \ref{fig:gammaDielectric} for the $\alpha$, $\beta$, and 
$\gamma$ phases, respectively, along with the real parts of the dielectric function $\epsilon_1(\omega)$ and the absorption coefficients $\alpha(\omega)$. The real parts are obtained from the imaginary parts via a Kramers-Kronig transformation, and the optical absorption from the relation $\alpha(\omega) = 2 \epsilon_2(\omega) / n(\omega)$, where $n(\omega)$ is the real part of the index of refraction $\tilde n(\omega) = \sqrt{\epsilon_1(\omega) + i \epsilon_2(\omega)}$. The onsets of absorption and their respective
polarizations are consistent with the symmetry analysis in the previous section. For $\alpha$, the top VBM at $\Gamma$ is  $E$
like and hence has allowed transitions for $x,y$ polarizations, while 
the VBM-1 has $A_1$ symmetry and hence has allowed transitions for 
$z$-polarization. For $\beta$ the top VBM at $\Gamma$ is forbidden 
but very close to the $a_1$ symmetry level allowed for $z$, the next ones have $b_1$ corresponding to $x$ and $b_2$ corresponding to $y$ 
symmetries and hence the onset occur in the order $z,x,y$. 
For the $\gamma$-structure  the top valence band at $\Gamma$  transition to the CBM is  forbidden but transitions along $\Gamma-X$
become allowed for $z$-polarization and along $\Gamma-M$ for $x,y$
polarization. This is consistent with the calculated onsets of 
absorption. 
Because the matrix elements are expected to gradually increase as  ${\bf k}$ moves away from $\Gamma$ either along $\Gamma-X$ or $\Gamma-M$ the onset of optical transitions would not follow the usual $\sqrt{E-E_0}$ behavior for direct allowed transitions but rather a $(E-E_0)^{3/2}$ behavior, where $E_0$ 
is the onset of transitions. On the other hand, even for the direct 
allowed transitions in the $\beta$ case, the expected $\sqrt{E-E_0}$
behavior is only seen very close to the onset. This is because 
the conduction band at somewhat higher energies above the CBM becomes linear in ${\bf k}$, in which case, the $\epsilon_2(\omega)$ will 
turn over to become proportional to $\omega^2$. In the $\gamma$-phase
we have a combination of linearly increasing matrix elements and 
a conduction band that turns from parabolic to linear behavior but 
in any case, it is clear from the figures that the  absorption coefficient turns on slower for the $\gamma$ than for the $\beta$
case. 

At present, no experimental data on the optical absorption or dielectric function
over the range we have calculated, are available. This is not surprising giving the difficulty
to obtain adequate light sources 
in this deep UV range, which is only available at synchrotrons.
As mentioned already in the introduction, there were two prior experimental studies \cite{Huang2008,Zou2005}
which showed a decrease in transmission at about 190 nm or 6.5 eV for the $\gamma$-phase. No data
are available for the high-pressure phases.  These onsets  of absorption are significantly lower
than the here calculated quasiparticle band gaps.  There are two possible reasons for this discrepancy:
finite temperarature effects and zero-point motion of the band gap and secondly, excitonic effects. 
In a very ionic material, as we have here, excitonic effects can be expected to be large and could
substantially reduce the optical absorption onset from the band-to-band onset.  An initial estimate can be made within the hydrogenic Wannier exciton model. Using a reduced mass of 0.33 and a
dielectric constant of about 3.5, the binding energy would be 0.36 eV.
On the other hand, electron-phonon coupling band gap renormalization effects could also be substantial leading to a zero point motion correction and even larger reductions at room temperature. A full calculation
of these effects is beyond the scope of the present paper. 
However, comparing to other ionic oxides like MgO suggest the finite temperature effects could well be of order 0.5 eV. Combining this with the estimated exciton binding energy a reduction by about 0.9-1.0 eV would bring the absorption onset  down to 6.7 eV in reasonable agreement with experiment.
Finally, we should point out that the optical measurements carried out thus far had a cut-off at about 190 nm in the UV and only a start of the reduction of transmission was measured at this wave length.
The nature of the absorption onset, which we here 
predict to be forbidden direct is therefore still unclear but is challenging to measure with standard available light sources in the UV.
Furthermore, defect related absorption band tails can often lead to an underestimate of the band gap. 

\subsection{Silicon Doping of $\gamma$-$\mathrm{LiAlO}_2$}
With band gaps larger than $7$ eV, the utility of LiAlO$_2$ in ultra-wide band gap (UWBG) semiconductor applications will depend heavily on whether it can successfully  be doped. Otherwise, it is just another insulator.
The electrical conductivity is also of importance in the context of Li-diffusion and the opportunities for
LiAlO$_2$ as electrodes in  Li batteries. For successful $n$-type doping one must find a dopant which
leads to shallow donor levels in the gap and can be readily introduced in the material.
For example, Si is used as $n$-type dopant in GaN but in Al$_x$Ga$_{1-x}$N alloys of high Al content $x>0.8$
leads to a  deep donor due to a distortion of the defect structure away from the simple substitutional
cation site,\cite{Mehnke2013} which is called a DX type defect. Donor binding energies larger than a few $0.1$ eV are usually
considered ineffective for doping.  Inspired by prior work on LiGaO$_2$ \cite{Dabsamut2020}
we were consider Si doping in LiAlO$_2$.

In this section we present the results of a band structure calculation for $\gamma$-LiAlO$_2$ doped with silicon as a candidate $n$-type dopant.  First, we simply replaced one Al in the 16 atom 
unit cell by Si, which corresponds to 25 \% doping. While this is 
an unrealistically high doping level, the advantage is that we can readily perform the calculation at the QS$GW$ level. Next, we studied a 128 atom $2\times2\times2$ supercell, with a single Si$_\mathrm{Al}$
corresponding to 3.125\% of the Al cation sublattice. We first calculate it at the GGA level. Finally, we use a recently developed cut-and-paste approach \cite{Dernek2022} to obtain approximately the QS$GW$ band structure in 
the 128 atom cell with a single Si, from the real-space self-energy 
of the perfect crystal and the self-energy matrix of the Si atom 
and its neighbors from the 16 atom QS$GW$ calculation. 
In this approach we use a cut-off of the real space self-energy 
matrices of 6.27 \AA. 
\begin{figure}
    \centering
    \includegraphics[width=7cm]{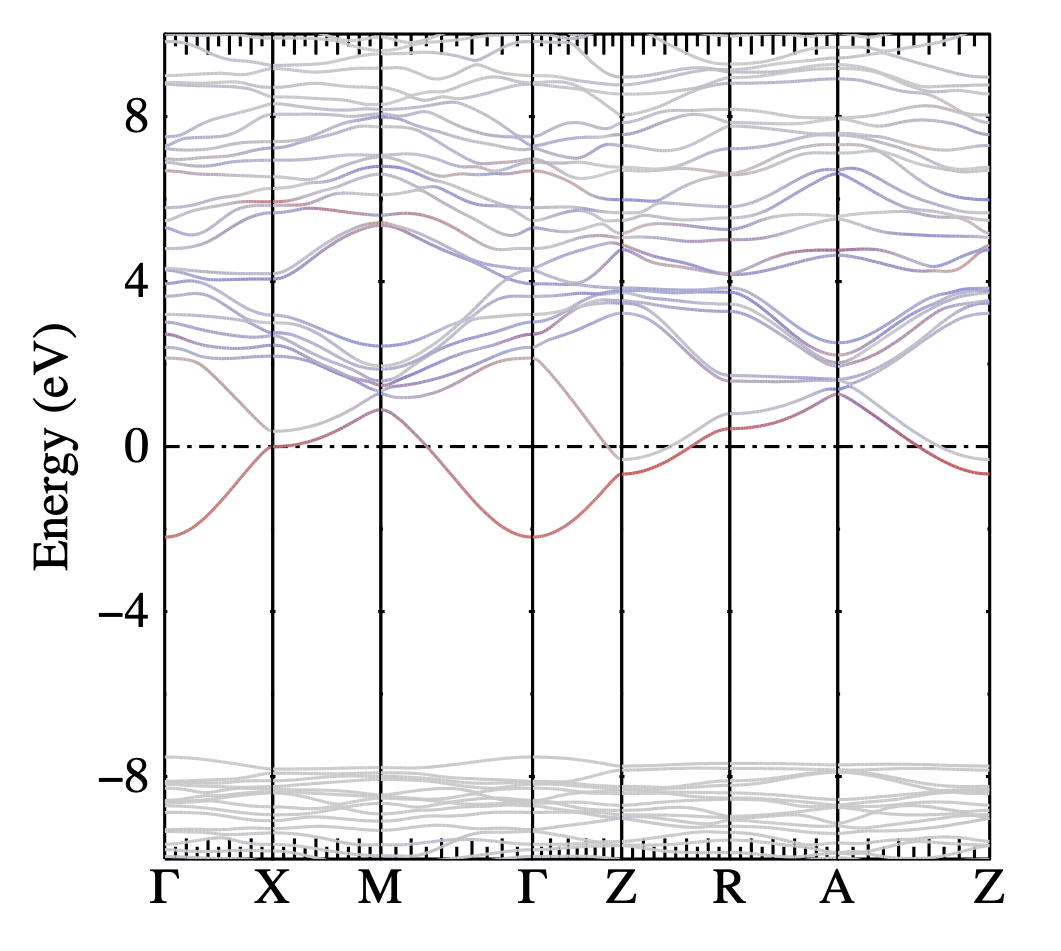}
    \caption{Band structure of SiAl$_3$O$_8$ in the $\gamma$-structure and in the QS$GW$ approximation. The red (blue) color indicates the Si-$s$ Si-$p$ orbital contributions with background bands in grey}
    \label{fig:sialo2}
\end{figure}
Fig. \ref{fig:sialo2} shows the results of the 16 atom cell. 
The Fermi level  now lies about 2 eV above  the CBM  indicating that no new levels occur in the gap
and the additional electron just starts to fill the conduction band. 
However, the bottom of the conduction band is clearly strongly Si dominated and the QS$GW$ gap at 5.33 eV is significantly lower than that of pure $\gamma$-LiAlO$_2$.  This might at first sight indicate that the Si may actually introduce a deep donor level, which is here broadened into an impurity band because of the high Si concentration and merging with the conduction band.   We need a larger cell to refute this possibility. 
\begin{figure}
    \centering
    \includegraphics[width=7cm]{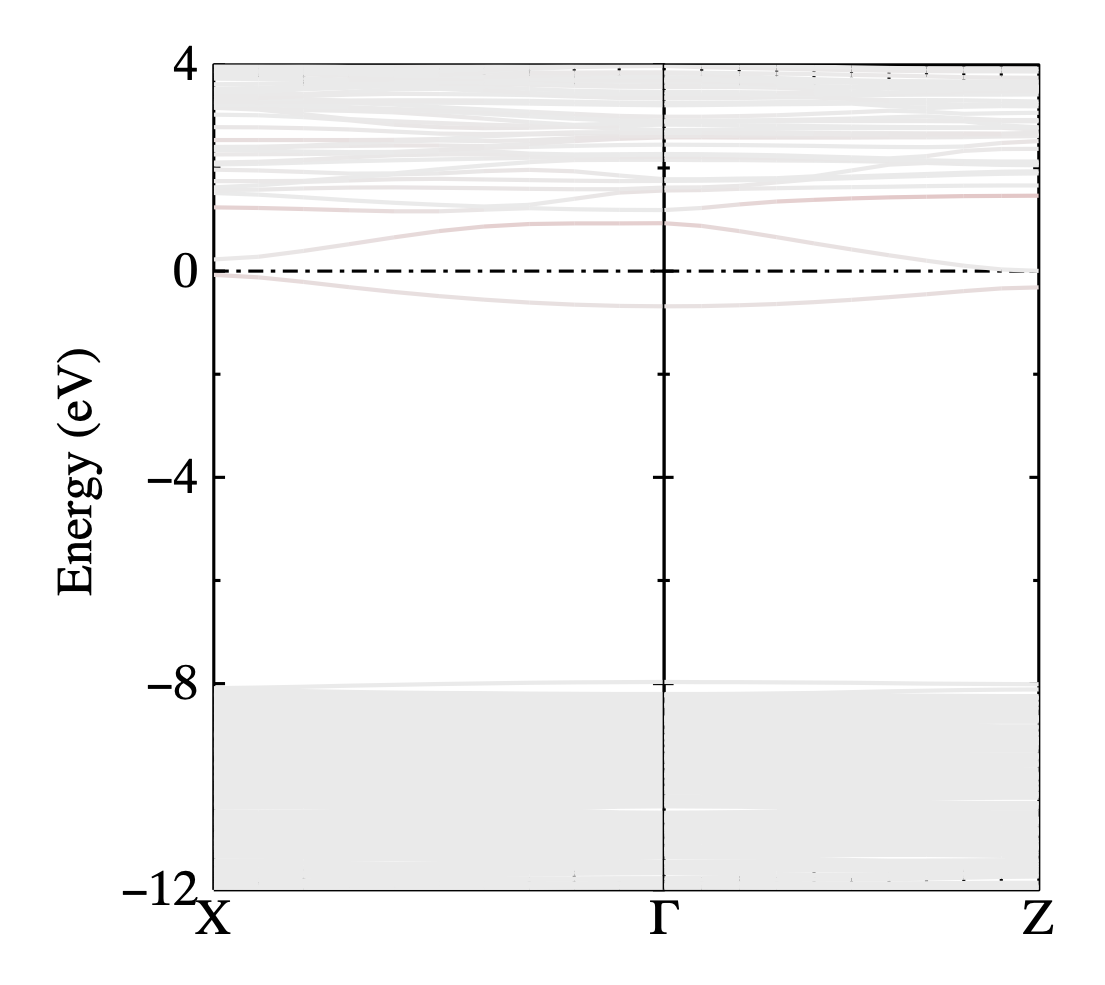}
    \caption{Band structure of $\gamma$-LiAlO$_2$:Si in a 128 atom supercell in the cut-and-paste QS$GW$ method. The red-color indicates Si-$s$ like orbital contribution.}
    \label{fig:lialo2_si}
\end{figure}

The results for the 128 atom cell are shown in Fig. \ref{fig:lialo2_si}.
We here show only  two directions of the Brillouin zone.
The faint red-color of the lowest three conduction bands indicates their
small Si-$s$ contribution.  The back ground color for the bands without any Si-$s$ was chosen a light grey because otherwise it would obscure  the red color.
The lowest three conduction bands both have some Si-$s$ contribution but it is less pronounced because the Si concentration is lower and the Fermi level now lies only 0.67 eV above the CBM. These  bands essentially are the folded version of the conduction band of the 16 atom cell in the Brillouin zone of the supercell, which has half the size in each direction, although at the new Brillouin zone edge, a gap opens. The band gap between the O-2p like VBM and the Si-like CBM  is about 7.28 eV and is close that of the corresponding perfect crystal gap of 7.43 eV. It is here a little bit lower than the converged value of Table
\ref{tab:gaps} because of the real-space cut-off of the self-energy. Thus the Si still pushes the CBM slightly down but less so than in the 16 atom cell and there is no evidence 
of a deep donor level. Compared to the corresponding GGA results, 
the CBM is simply pushed up along with its Si contribution. 
These results did not include relaxation of the structure. 
However, relaxation of the nearest neighbor O atoms around the Si, 
show an inward relaxation of the Si-O bonds by about 6 \%. They do not indicate a strong distortion or
DX center formation.  Indeed, it did not lead to any notable difference in the band structure in the GGA, which still does not show a defect level 
to emerge in the gap.

These results  indicate $n$-type doping should be possible.  However, 
 the hydrogenic donor model would predict $E_B=m_c^*R/\varepsilon^2$, which,  with $R$ the Rydberg unit (13.6057 eV), $m_c^*\approx0.4$
and $\varepsilon\approx3.5$, gives $E_B\approx0.4$ eV. This estimate includes only electronic screening. Estimating the phonon contributions to the screening requires the calculation of LO-phonons, which we have not yet done. However, comparison with LiGaO$_2$ indicates a static dielectric constant of about 6.5 is expected.  This would reduce the donor binding energy to about 0.1 eV. 
This shows that even if Si doping does not produce a well separated  defect band in the gap in a first-principles calculation  in the cell size we can here accommodate, it will likely act as a relatively deep donor with of order 100 meV binding energy. This is similar to LiGaO$_2$ for which experimental confirmation of $n$-type doping 
also is still to be accomplished but promising enough to warrant experimental attempts
to dope these materials with Si. While these estimates indicate 
efficient $n$-type doping may be challenging for both LiGaO$_2$ 
and LiAlO$_2$, they would significantly extend the gap range of 
UWBG semiconductors. Some compromise between efficiency of doping and
larger band gaps may be necessary. For comparison, in $\beta$-Ga$_2$O$_3$, the 
conduction band effective mass is about 0.3 and the dielectric constant about 10, giving a binding energy of order 40 meV.
The expected donor binding energy in LiGaO$_2$ and LiAlO$_2$ are thus 
relatively high, but on the other hand, still smaller than typical acceptor binding energies related to $p$-type doping in for example GaN.

\section{Conclusions}
The main conclusions of this paper are as follows. First, in terms of
structures, 
the $\gamma$-phase has the lowest energy but is very close to the
$\beta$-phase, which may already be stabilized at slight pressures of about 0.2 GPA.  A phase transition from these tetrahedrally bonded phases to the octahedrally bonded $\alpha$-phase is predicted to occur near 1GPa. The band structures of the three phases were obtained in the QS$GW$ method and 
yield band gaps larger than 7 eV with a pseudodirect gap of 7.69 eV in
the $\gamma$-phase, indirect gap of 8.16 eV in the $\beta$-phase, and indirect gap of
9.30 eV in the $\alpha$-phase.  The calculated onset of 
absorption in the $\gamma$-phase, slightly above the pseudodirect gap 
of 7.7 eV and stemming from direct transitions along $\Gamma-M$ and 
$\Gamma-X$ is significantly higher than the until now reported optical absorption onset of about 6.5 eV.
We suggested this can be explained by a combination of finite temperature renormalization effects on the gap
due to electron-phonon coupling and  strong excitonic effects in this ionic material. A fuller
investigation of these effects will require additional work in the future. 
For future use, we have provided details of the band structure 
near the band edges, including the effective masses. We have also briefly discussed the possibility of silicon as a donor for $\gamma$-LiAlO$_2$, which indicates that $n$-type doping should be possible although efficient doping will be challenging due to the relatively high donor binding energy. 
\acknowledgements{
This work made use of the High Performance Computing Resource in the Core Facility for Advanced Research Computing at Case Western Reserve University. This work was supported by the U.S. Department of Energy-Basic Energy Sciences  
under grant No. DE-SC0008933 and a student travel grant from CWRU SOURCE.}

\nocite{*}

\bibliography{Bibliography}

\end{document}